\documentclass[twocolumn]{autart}

\usepackage{graphicx}          

\usepackage{natbib}        

\usepackage{amsmath}
\usepackage{amssymb}
\usepackage{mathtools, cuted}
\usepackage{lipsum}
\def\endprf{{$\hfill\square$}}

\expandafter\def\expandafter\normalsize\expandafter{%
    \normalsize
    \setlength\abovedisplayskip{2pt}
    \setlength\belowdisplayskip{2pt}
}

\begin{document}

\begin{frontmatter}

\title{Feedback Passivation of Linear Systems with Fixed-Structured Controllers\thanksref{footnoteinfo}}

\thanks[footnoteinfo]{This paper was not presented at any IFAC 
meeting. }

\author[1]{Lanlan Su}\ead{lanlansu.work@gmail.com},    
\author[1]{Vijay Gupta}\ead{vgupta2@nd.edu}, and   
\author[1]{Panos Antsaklis}\ead{antsaklis.1@nd.edu }       

\address[1]{Department of Electrical Engineering,
University of Notre Dame, Notre Dame, IN 46556 USA}  

\begin{keyword}
{Output feedback passivation}, 
{Fixed-structued controller},
{Passivity indices},
{SDP}.

\end{keyword}                            

\begin{abstract}  
This paper addresses the problem of designing an optimal output feedback controller with a specified controller structure for linear time-invariant (LTI) systems to maximize the  passivity level for the  closed-loop system, in both continuous-time (CT) and discrete-time (DT).   Specifically, the set of controllers under consideration is linearly parameterized with constrained parameters. Both input feedforward passivity (IFP) and output feedback passivity (OFP) indices are used to capture the level of passivity. Given a set of stabilizing controllers,  
 a necessary and sufficient condition is proposed for the existence of such fixed-structured output feedback controllers  that can passivate the closed-loop system. Moreover, it is shown that the condition can be used to obtain the controller that maximizes the IFP or the OFP index  by solving a convex optimization problem. 
\end{abstract}

\end{frontmatter}

\section{Introduction}

Passivity provides a physically meaningful interpretation of the energy dissipation of a system from the input-output perspective (\cite{willems1972dissipative}). Notions of input and output passivity indices give a widely used measure of the level of passivity for a system (\cite{kottenstette2014relationships}). When exploited properly, passivity indices provide a means to design feedback controllers  via the process of compensating for the lack of passivity in one subsystem of a feedback configuration with passivity surplus in the other \citep{van2000l2,bao2007process,antsaklis2013control}.

Feedback passivation of plants that may not be passive is a widely studied problem (\cite{larsen2001passivation, zhu2014passivity,zhao2016feedback}). In the existing works, the controller can be chosen without any constraint on its structure.  In this paper, we study  the problem of designing  a controller to maximize the closed-loop passivity level (as measured by a passivity index) when the controller has to satisfy a fixed structure.   Reconstruction of such a controller can be performed by solving a semidefinite programming (SDP). The results in this paper can be utilized to improve the robust stability margins of interconnected systems as measured from the perspective of passivity.

Our problem setup involves linearly parameterized sets of controllers with constrained parameters. It is generally known that synthesis of such controllers is a notoriously difficult problem. A naive application of the Kalman-Yakubovich-Popov (KYP) lemma to tackling the problem would result in bilinear matrix inequalities, which are intractable in general. Our proposed approach establishes and exploits relationships between the passivity of SISO systems and sum-of-square (SOS) polynomials, which are  amenable to convex optimization formulations. The introduction of linearly parameterized sets of controllers is motivated by its ubiquity in practical engineering applications, the most common of which consists of proportional-integral-derivative (PID) controllers, where the parameters appear linearly. By tuning the linear parameters of the controller in order to optimize the level of passivity in a feedback is of interest in view of the popularity of such controllers. Such parameterized controllers belong to a broader class of the so-called fixed-structured controllers. See, for instance, \cite{saeki2006fixed} which considers fixed-structured PID controller design  for $H_\infty$ control problems with  linear constraints on the control structure; \cite{malik2008linear} wherein  a set of stabilizing 
fixed-structure and fixed-order controllers is constructed; and \cite{bazanella2011data} which studies model-free fixed-structure controller synthesis. To the best of the authors' knowledge, the problem of optimizing the passivity level of a system with a fixed-structure controller considered in this paper has not been considered elsewhere.

 The rest of this paper is organized as follows.   Section 2 introduces some preliminaries and states the problem formulation. Section 3 presents the main results.  The main results are illustrated by two examples in Section IV. Some final remarks and future work are described in Section V.

\section{Preliminaries and Problem Formulation}

\subsection{Notation}
The notation used in the paper is as follows. The sets of real and complex numbers are denoted by $\mathbb{R}$ and $\mathbb{C}$, and the imaginary unit is denoted as $j$. The notation $Re(\lambda)$ and $|\lambda|$ denote the real part and the magnitude of a complex number $\lambda$.   $A^{-1}$, $A^{'}$ and $A^{*}$ denote the inverse, the transpose and the conjugate transpose of matrix $A$, respectively. Given a Hermitian matrix $A=A^{*}$, the notation $\underline{\lambda}(A)$ denotes the  minimum eigenvalue of $A$. For Hermitian matrices $A,B$, the notation $A-B\ge0$ denotes $A-B$ is positive semidefinite. 
 The degree of a polynomial $p(\cdot)$ is denoted by $\text{deg}(p(\cdot))$. The symbol $\otimes$ denotes the Kronecker product. 
 The function $f_T$ is the truncation of $f$ to the interval $[0,T]$. 
 The operator $<f,g>_T$ is defined as the inner products of signal $f$ and $g$ over $[0,T]$. $\mathcal{L}_{2e}$ denotes the extended $L_2$ signal space and $||\cdot||$ denotes the $L_2$ norm. For briefness, the notation $\star$ denotes the symmetric entries in a symmetric matrix.

\subsection{Sum of square (SOS) matrix polynomial}\label{sos}

Let us briefly introduce the class of SOS matrix polynomials, see
e.g., \cite{chesi2010lmi} for
details.

A symmetric matrix polynomial $F:\mathbb{R}^{r}\rightarrow\mathbb{R}^{n\times n}$
is said to be SOS if and only if there exist matrix polynomials $F_{1},\ldots,F_{k}:\mathbb{R}^{r}\rightarrow\mathbb{R}^{n\times n}$
such that $F(s)=\sum_{i=1}^{k}F_{i}(s)^{T}F_{i}(s).$ SOS matrix polynomials
are postive-semidefinite, and it turns out that one can establish
whether a symmetric matrix polynomial is SOS via an LMI feasibility
test.

Indeed, let $d$ be a nonnegative integer such that $2d\ge\deg(F)$.
By extending the Gram matrix method or SMR for scalar polynomials
to the representation of matrix polynomials, $F(s)$ can be written
as 
\begin{equation}
F(s)=\left(b(s)\otimes I\right)^{T}\left(M+L(\alpha)\right)\left(b(s)\otimes I\right)\label{smrmat}
\end{equation}
where $b(s):\mathbb{R}^{r}\rightarrow\mathbb{R}^{\sigma(r,d)}$ is
a vector containing all the monomials of degree less than or equal
to $d$ in $s$ with $\sigma(r,d)=\frac{(r+d)!}{r!d!}$ , and $M\in\mathbb{R}^{n\sigma(r,d)\times n\sigma(r,d)}$
is a symmetric matrix satisfying 
\[
F(s)=\left(b(s)\otimes I\right)^{T}M\left(b(s)\otimes I\right),
\]
 $L(\alpha):\mathbb{R}^{\omega(r,2d,n)}\rightarrow\mathbb{R}^{n\sigma(r,d)\times n\sigma(r,d)}$
is a linear parametrization of the linear set 
\begin{equation}
\mathcal{L}=\{\tilde{L}=\tilde{L}^{T}:~\left(b(s)\otimes I\right)^{T}\tilde{L}\left(b(s)\otimes I\right)=0\},\label{L}
\end{equation}
and $\alpha\in\mathbb{R}^{\omega(r,2d,n)}$ is a free vector with
$\omega(r,2d,n)=\frac{1}{2}n\left(\sigma(r,d)(n\sigma(r,d)+1)-(n+1)\sigma(r,2d)\right).$
It follows that $F(s)$ is a SOS matrix polynomial if and only if
there exists $\alpha$ satisfying the LMI 
\[
M+L(\alpha)\ge0.
\]
When $n=1$ is considered, the above results are reduced to  SOS
polynomials.

\subsection{Strictly Input and Output Passive Systems}
We start by introducing the definitions of passivity and positive realness for LTI systems, followed by a lemma revealing their relation. 
\begin{defn}
(Passivity~\citep{van2000l2,kottenstette2014relationships}) Consider a CT or DT LTI system $H:\ u\in\mathcal{L}_{2e}\rightarrow y\in\mathcal{L}_{2e}$.
Then the system $H$ is
\begin{itemize}
\item passive if there exists a constant $\beta$ such that 
\begin{equation}
\left\langle Hu,u\right\rangle _{T}\ge\beta,\forall u\in\mathcal{L}_{2e},\forall T\ge0.\label{eq:def: passive}
\end{equation}
\item strictly input passive (SIP) if there exist $\nu>0$ and $\beta$
such that 
\begin{equation}
\left\langle Hu,u\right\rangle _{T}\ge\nu||u_{T}||_{2}^{2}+\beta,\forall u\in\mathcal{L}_{2e},\forall T\ge0,\label{eq:def nu}
\end{equation}
and the largest $\nu>0$ satisfying \eqref{eq:def nu} is called the
Input Feedforward Passivity (IFP) index, denoted as IFP($\nu$). 
\item strictly output passive (SOP) if there exist $\xi>0$ and $\beta$
such that 
\begin{equation}
\left\langle Hu,u\right\rangle _{T}\ge\xi||Hu_{T}||_{2}^{2}+\beta,\forall u\in\mathcal{L}_{2e},\forall T\ge0,\label{eq:def xi}
\end{equation}
and the largest $\xi>0$ satisfying \eqref{eq:def xi} is called the
Output Feedback Passivity (OFP) index, denoted as OFP($\xi$). 
\end{itemize}
\end{defn}

The IFP and OFP indices, defined in terms of an excess of passivty, are introduced
to quantify the degree of passivity. 

\begin{defn}
\label{def:Positive-realness}(Positive realness~\citep{khalil2002nonlinear}) A square, proper and rational transfer
function $G(s)$ (or $G(z)$ for DT case) is said to be positive real
if 
\begin{itemize}
\item $G(s)$ is analytic in $\text{Re}(s)>0$ in CT case; $G(z)$ is analytic
in $|z|>1$ in DT case;
\item $G(jw)+G^{*}(jw)\ge0$ , $\forall\omega\in\mathbb{R}$ for which $j\omega $ is not a pole of $G(s)$ in CT case;
$G(e^{j\omega})+G^{*}(e^{j\omega})>0,\forall\omega\in[0,2\pi]$ for which $e^{j\omega}$ is not a pole of $G(z)$ in
DT case;
\item Any pure imaginary pole $j\omega_{o}$ of $G(s)$ is a simple pole,
and the associated residue $G_{o}\triangleq \lim_{s\rightarrow j\omega_{o}}(s-j\omega_{o})G(s)$
satisfies $G_{o}=G_{o}^{*}\ge0$ in CT case; If $e^{j\omega_{o}}$
is a pole of $G(z)$ it is at most a simple pole, and the associated
residue $G_{o}\triangleq\lim_{z\rightarrow j\omega_{o}}(z-e^{j\omega_{o}})G(s)$
satisfies $G_{o}=G_{o}^{*}\ge0$ in DT case. 
\end{itemize}
\end{defn}

For a stable\footnote{In this work, a LTI system is said to be stable if the system is asymptotically stable.} LTI system with transfer function 
$G$, the following lemma states the relation between the passivity and
positive realness.
\begin{lem}
\label{lem:relation}(\cite{bao2007process})A stable LTI system $H:\ u\in\mathcal{L}_{2e}\rightarrow y\in\mathcal{L}_{2e}$
is passive if and only if its transfer function $G$ is positive real. 

For a stable LTI system with the transfer function $G(s)$
(or $G(z)$ for DT case) that is strictly input passive, its IFP index,
$\nu$, is given as
\[
\nu=\left\{ \begin{array}{ll}
\frac{1}{2}\underset{\omega\in\mathbb{R}}{\text{min}}\;\;\underline{\lambda}\left(G(j\omega)+G^{*}(j\omega)\right) & CT\ case\\
\frac{1}{2}\underset{\omega\in[0,2\pi]}{\text{min}}\underline{\lambda}\left(G(e^{j\omega})+G^{*}(e^{j\omega})\right) & DT\ case
\end{array}\right.
 \]
 
 For a minimum phase  LTI system $G(s)$ (or $G(z)$ for DT case) that is strictly output passive, its OFP index, $\xi$, is given as   
 \[
\xi=\left\{ \begin{array}{ll}
\frac{1}{2}\underset{\omega\in\mathbb{R}}{\text{min}}\;\;\underline{\lambda}\left(G^{-1}(j\omega)+[G^{-1}(j\omega)]^{*}\right) & CT\ case\\
\frac{1}{2}\underset{\omega\in[0,2\pi]}{\text{min}}\underline{\lambda}\left(G^{-1}(e^{j\omega})+[G^{-1}(e^{j\omega})]^{*}\right) & DT\ case
\end{array}\right.
 \]
\end{lem}

\subsection{Problem Formulation}
\begin{figure}
\centering
 \includegraphics[scale = 0.7]{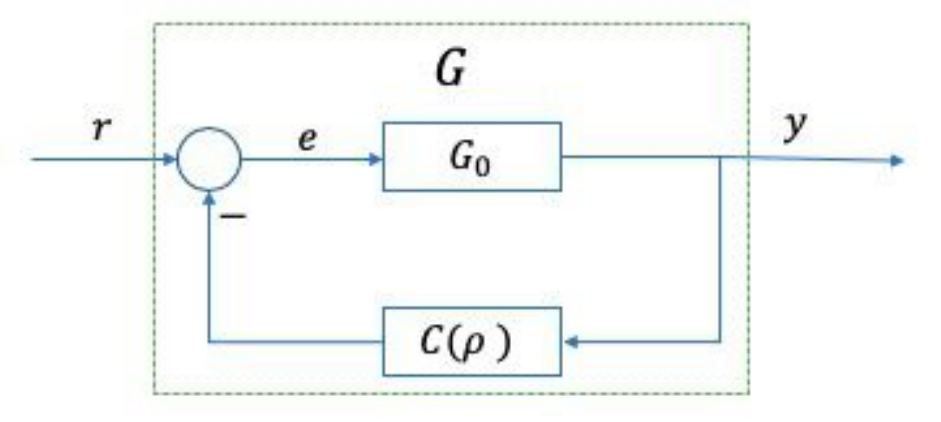}
 \caption{Feedback control system}\label{fig:1}
\end{figure}

In this work, we consider the problem of output feedback passivation of a single-input single-output (SISO) linear system through a fixed-structured controller (depicted in Figure \ref{fig:1}). Particularly, the objective is to design an output feedback fixed-structure controller with  parameter $\rho^{*}$, which maximizes the IFP or the OFP index for the closed-loop system. 

The SISO plant with transfer function $G_0$ can be either CT or DT systems.  The set of  the controllers which can be implemented has a specified controller structure  represented by $\mathcal{C}=\{C(s,\rho):\rho\in\mathcal{P}\}$
for the CT case and $\mathcal{C}=\{C(z,\rho):\rho\in\mathcal{P}\}$
for the DT case, where $\mathcal{P}\subseteq\mathbb{R}^{p}$ is a
set of admissible values of the controller parameter vector $\rho$. 
We assume that the controllers are linearly parameterized, i.e., 
\begin{equation}
C(s,\rho)=\rho^{T}\bar{C}(s),\ C(z,\rho)=\rho^{T}\bar{C}(z)\label{eq:linear paramterization}
\end{equation}
where $\rho$ is the parameter vector and $\bar{C}$ is the predefined parameter independent
vector of transfer functions. It is also assumed that all entries in $\bar{C}$  are selected to have stable poles.  A typical class of controllers with linear
parameterization is PID controllers. The linearity makes the
resulting design problem more amendable  to analysis. Moreover, it is
shown in \cite{bazanella2011data} that any parameteter-dependent transfer function
can be approximated to any degree of accuracy desired by a transfer
function of the form \eqref{eq:linear paramterization} with sufficiently
large $p$. As it is often required to restrict the admissible controller
parameters to some desired bounded sets, we assume that the admissible
set of $\rho$ is described by\footnote{As it will be explained in Remark \ref{rem:footnote}, the proposed methodology can be used
also to design feedback controllers with any convex set $\mathcal{P}$.} 
\begin{equation}
\mathcal{P}=\{\rho\in\mathbb{R}^{p}:\underline{\rho}_{i}\le\rho_{i}\le\overline{\rho}_{i},i=1,\ldots,p\}.\label{eq:P set}
\end{equation}
The problems addressed in this work are as follows.
\begin{prob}
\label{prob:1}For a given set of controllers, $\mathcal{C}=\{C(\rho):\rho\in\mathcal{P}\}$,
establish whether the closed-loop system is stable for all $\rho\in\mathcal{P}$.
\end{prob}
With the set  of stabilizing controllers in hand, we further investigate the following problem. 
\begin{prob}
\label{prob:2}Establish whether there exists a controller $C$ in
the set $\mathcal{C}$ that can passivate the system $G_0$.
If the answer is positive, determine the controller $C^{*}$ that
maximizes the  IFP index  and the OFP index  respectively for the closed-loop system. 
\end{prob}

It is well-known that a necessary condition for a linear system to be feedback passivated is that the system should
have a relative degree less than 2 and is weakly minimum phase (i.e., it should not have zeros on right side in s-plane or  outside the unit circle in z-plane). Thus, we assume throughout this work  the following assumption.

\begin{assum}\label{assum:1}
The plant $G_0$ has a relative degree less than $2$, and has all its zeros in the closed left half of the s-plane  in CT case (in DT case, respectively, inside or on  the unit circle of the z-plane).  
\end{assum}
A slightly more restrictive assumption is made when the optimal OFP controller design is considered. 
\begin{assum}\label{assum:2}
The plant $G_0$ has a relative degree less than $2$, and has all its zeros in the open left half of the s-plane  in CT case (in DT case, respectively, strictly inside  the unit circle of the z-plane).  
\end{assum}

\section{Main Results}
\subsection{Stability Analysis}

Let us start by addressing Problem \ref{prob:1}, which is to establish the robust stability of the closed-loop system for all parameter $\rho\in\mathcal{P}$. 

First, let us observe that the controller set \eqref{eq:P set} can
be equivalently described as 
\begin{equation}
\mathcal{P}=\{\rho\in\mathbb{R}^{p}:c_{i}\ge0,i=1,\ldots,p\}\label{eq:P set modified}
\end{equation}
with $c_{i}=(\overline{\rho}_{i}-\rho)(\rho_{i}-\underline{\rho}_{i})$. 

For CT case, let us denote the transfer function of the plant
as $G_{0}(s)=\frac{N_{0}(s)}{D_{0}(s)}$, and denote the $i$-th component
in the vector $\bar{C}(s)$ as $\bar{C}_{i}(s)=\frac{N_{i}(s)}{D_{i}(s)}$.
It follows that the closed-loop system as shown in Figure 1 is represented
as
\begin{equation}
\begin{array}{cl}\vspace{1mm}
 G(s,\rho) & =\displaystyle\frac{G_0(s)}{1+G_0(s)C(s,\rho)}\\
\vspace{1mm}
= & \frac{N_{0}(s)\prod_{i=1}^{p}D_{i}(s)}{D_{0}(s)\prod_{i=1}^{p}D_{i}(s)+N_{0}(s)\sum_{i=1}^{p}\rho_{i}N_{i}(s)\left(\prod_{j\neq i}D_{j}(s)\right)}\\
\triangleq & \displaystyle\frac{p_N(s)}{p_D(s,\rho)}
\end{array}\label{eq:closed-loop transfer 1}
\end{equation}
where the polynomials $p_N(s)$ and $p_D(s,\rho)$ denote the numerator and denominator of the closed-loop transfer function respectively. Under Assumption \ref{assum:1} or \ref{assum:2} , it can be observed that
there is no unstable zero-pole cancellation in the above closed-loop
transfer function. 

Rewrite the denominator polynomial $p_D(s,\rho)$ as 
\[
p_D(s,\rho)=a_{n}(\rho)s^{n}+a_{n-1}(\rho)s^{n-1}+\ldots+a_{1}(\rho)s+a_{0}(\rho)
\]
wherein the coefficients $a_{0},\ldots,a_{n}$ are linear functions of the vector variable $\rho$. 
In order to analyze the stability of the closed-loop system, it is necessary and sufficient to check whether
all the roots of  the polynomial $p_D(s,\rho)$ have negative
real parts for all $\rho\in\mathcal{P}$. To this end, let us exploit the modified Routh-Hurwitz
table for the polynomial $p_D(s,\rho)$. By multiplying each component
by their denominator in the classical Routh-Hurwitz table, we can
obtain the modified Routh-Hurwitz table defined as 
\begin{equation}
\begin{array}{cccc}
a_{n}(\rho) & a_{n-2}(\rho) & a_{n-4}(\rho) & \cdots\\
a_{n-1}(\rho) & a_{n-3}(\rho) & a_{n-5}(\rho) & \cdots\\
a_{31}(\rho) & a_{32}(\rho) & a_{33}(\rho) & \cdots\\
\vdots & \vdots & \vdots & \ddots
\end{array}\label{M_RH TABLE}
\end{equation}
 where the number of rows is $n+1$ and the $ij$-th component is 
\begin{equation}
\begin{array}{c}
a_{ij}(\rho)=a_{i-1,1}(\rho)a_{i-2,j+1}(\rho)-a_{i-1,j+1}(\rho)a_{i-2,1}(\rho)\\
\;i=3,\ldots,n+1,j=1,2,\ldots
\end{array}\label{eq:M_RH}
\end{equation}
 It can be verified that all the roots of $p_D(s,\rho)$ have
negative real parts for all $\rho\in\mathcal{P}$ if and only if the
polynomials in the first column of the modified Routh-Hurwitz table
\eqref{M_RH TABLE} are positive for all $\rho\in\mathcal{P}$. 

Now let us further consider a discrete-time transfer function of the
plant as $G_{0}(z)=\frac{N_{0}(z)}{D_{0}(z)}$, which is in closed-loop
with the linearly parameterized controller $C(z,\rho)=\sum_{i=1}^{p}\rho_{i}\frac{N_{i}(z)}{D_{i}(z)}$.
With similar argument of the CT case, the closed-loop system
is represented as 
\begin{equation}
G(z,\rho) = \frac{p_N(z)}{p_D(z,\rho)}\label{eq:close-loop system_DT}
\end{equation}
with
\[
p_D(z,\rho)=a_{n}(\rho)z^{n}+a_{n-1}(\rho)z^{n-1}+\ldots+a_{1}(\rho)z+a_{0}(\rho)
\]
wherein the coefficients $a_{0},\ldots,a_{n}$ depend linearly on
the vector variable $\rho$. Similarly, in order to establish the stability of the closed-loop system for all $\rho\in\mathcal{P}$, it is necessary and sufficient
to check whether all the roots of  the polynomial $p_D(z,\rho)$
have magnitude less than 1.   By multiplying the odd rows by their denominator
and removing the even rows in the traditional Jury table, we define
the  modified Jury table
where
\[
\begin{array}{rl}
 & a_{ij}(\rho)\\=&a_{i-1,j}(\rho)a_{i-1,1}(\rho)-a_{i-1,n+4-i-j}(\rho)a_{i-1,n+3-i}(\rho)\\&
i=3,\ldots,n+1,\:j=1,2\ldots
\end{array}
\]
It can be verified that all the roots of $p_D(z,\rho)$ have magnitude
less than 1 for all $\rho\in\mathcal{P}$ if and only if all the polynomials
in the first column of the modified Jury table 
are positive for all $\rho\in\mathcal{P}$.

Let us denote the entries of the first column in the modified Routh-Hurwitz table for CT
case or in the modified Jury table for DT case as $f_{i}(\rho),i=1,\ldots,n+1$
where $f_{i}(\rho)$ denotes the $i$-th entry in the column. Based on the previous analysis, we have the following lemma. 

\begin{lem}
The closed-loop system is stable for all $\rho\in\mathcal{P}$ defined
in \eqref{eq:P set modified} if and only if 
\begin{equation}
f_{i}(\rho)>0,i=1,\ldots,n+1\ \forall\rho\in\mathcal{P}.\label{eq:stability cond 1}
\end{equation}
\end{lem}

Now let us define the polynomials
\begin{equation}
g_{i}(\rho)=f_{i}(\rho)-\sum_{j=1}^{p}s_{ij}(\rho)c_{j}(\rho)\ i=1,\ldots,n+1\label{eq:polynomial g}
\end{equation}
where $s_{ij}(\rho)$ are auxiliary polynomials.

\begin{thm}\label{thm:stability sos}The closed-loop system is stable for all $\rho\in\mathcal{P}$  if and only if 
\begin{equation}
\theta^{*}>0\label{eq:stability cond_2_1}
\end{equation}
where 
\begin{equation}
\theta^{*}=\underset{\theta,s_{ij}}{\text{max}}\;\;\theta\ \text{s.t.}\ \left\{ \begin{array}{rcl}
g_{i}(\rho)-\theta & \text{is} & \text{SOS}\\
s_{ij}(\rho) & \text{is} & \text{SOS}\\
\forall i & = & 1,\ldots,n+1\\
\forall j & = & 1,\ldots,p
\end{array}\right.\label{eq:stability_cond_2_2}
\end{equation}
\end{thm}
\textbf{Proof.}
$"\Rightarrow"$ It can be observed that the set $\mathcal{P}$ defined
in \eqref{eq:P set modified} is a compact set, and $c_{1},\ldots c_{p}$
are polynomials of even degree and their highest degree forms do not
have common zeros except zero. Given an arbitrarily small scalar $\theta>0$, it follows from Theorem 7 in \cite{chesi2010lmi} that $f_{i}(\rho)>\theta,\ \forall\rho\in\mathcal{P}$ holds if
and only if there exist SOS polynomials $s_{ij}(\rho)$ such that
$g_{i}(\rho)-\theta$ is SOS polynomial. Therefore, the condition \eqref{eq:stability_cond_2_2}
is satisfied with $\theta>0$, and hence the condition \eqref{eq:stability cond_2_1}
holds.

$"\Leftarrow"$ Let us suppose that \eqref{eq:stability cond_2_1}-\eqref{eq:stability_cond_2_2}
hold. Then, one has that $g_{i}(\rho)-\theta$ and $s_{ij}(\rho)$ are nonnegative. Since $c_j(\rho)\ge 0$ whenever $\rho\in\mathcal{P}$, it follows from \eqref{eq:polynomial g} that $f_{i}(\rho)>0,i=1,\dots n+1$ for all $\rho\in\mathcal{P}$.
\endprf
\vspace{1mm}
\begin{rem}
 Theorem \ref{thm:stability sos} shows that one can establish the positivity of the
polynomials in the first column in the modified tables  for
all $\rho\in\mathcal{P}$ by solving the optimization problem \eqref{eq:stability_cond_2_2}. It is worth mentioning that the condition for polynomials which depend on some decision variables linearly to be SOS polynomials can be solved
equivalently via LMIs based on the Gram matrix method as described in Section \ref{sos}. Therefore, for any chosen degrees of polynomials $s_{ij}(\rho)$, this theorem provides a sufficient
condition solvable through LMIs, which is also necessary when the degrees are large enough.
\end{rem}

Since fixed-structured controllers, including PID control as a typical example, are so widely used in industrial applications, it is important to develop a methodology to characterize the set of stabilizing controllers before carrying out the optimal control design. Theorem \ref{thm:stability sos} provides a method to establish  whether a given set of controllers is stabilizing. In the next subsection, we will design the controller by choosing its parameter $\rho$ from the set $\mathcal{P}$ to reach the maximized passivity level for the closed-loop system. Indeed, most existing modern optimal control techniques are incapable of accommodating constraints on the controller order
or structure into their design methods, and consequently cannot be used for designing optimal or robust controllers.

\subsection{Feedback Passivation}
Given a set of stabilizing controllers $\mathcal{C}$,
we proceed to address Problem \ref{prob:2} in this subsection. We first consider the CT case, which is then extended to the DT case. 

\subsubsection{CT case}
Recall that the numerator and denominator of the closed-loop transfer function \eqref{eq:closed-loop transfer 1} are denoted as  polynomials $p_{N}(s)$
and $p_{D}(s)$, respectively, wherein the coefficients of $p_{D}(s)$
depends linearly on the vector variable $\rho$. By substituting $s=j\omega$,
$p_{N}$ and $p_{D}$ can be rewritten via even-odd decomposition as 
\begin{equation}
\begin{array}{rcl}
p_{N}(j\omega) & = & p_{N}^{e}(w)+jp_{N}^{o}(w)\\
p_{D}(j\omega,\rho) & = & p_{D}^{e}(\omega,\rho)+jp_{D}^{o}(\omega,\rho)
\end{array}
\end{equation}
where $p_{N}^{e},p_{N}^{o},p_{D}^{e},p_{D}^{o}$ are all real polynomials
in $\omega$, and $p_{D}^{e}(\omega,\rho)$ and $p_{D}^{o}(\omega,\rho)$
depend linearly on $\rho$. The frequency response of the closed-loop system \eqref{eq:closed-loop transfer 1} can be expressed as 
\begin{equation}
G(j\omega,\rho)=\frac{p_{N}^{e}(w)+jp_{N}^{o}(w)}{p_{D}^{e}(\omega,\rho)+jp_{D}^{o}(\omega,\rho)}\label{eq:CT_closed-loop jw}
\end{equation}
which yields that 
\begin{equation}
\begin{array}{rl}
     &  G(j\omega,\rho)+G^{*}(j\omega,\rho)\\
     =& \displaystyle \frac{2p_{N}^{e}(w)p_{D}^{e}(\omega,\rho)+2p_{N}^{o}(w)p_{D}^{o}(\omega,\rho)}{p_{D}^{e}(\omega,\rho)^{2}+p_{D}^{o}(\omega,\rho)^{2}}.
\end{array}
\label{eq:closed-loop transfer 2}
\end{equation}

\begin{lem}\label{lm:3}
There exists a controller $C(s,\rho)$ in the controller set $\mathcal{C}$
that can feedback passivate the plant $G_{o}(s)$ if and only if there
exists a vector $\rho\in\mathcal{P}$ and a scalar $\epsilon\ge0$
such that 
\begin{equation}
p_{N}^{e}(w)p_{D}^{e}(\omega,\rho)+p_{N}^{o}(w)p_{D}^{o}(\omega,\rho)-\epsilon\text{ is SOS}.\label{eq:lmi cond 1}
\end{equation}
\end{lem}
\textbf{Proof.}
Since the controller set $\mathcal{C}$ is stabilizing, it follows from Lemma \ref{lem:relation} that a controller $C(s,\rho)$ can feedback passivate the plant  if and only if the closed-loop system is positive real. For a stable closed-loop system, the
first and the third condition in Definition \ref{def:Positive-realness}
are trivially satisfied. Therefore,
the closed-loop system is positive real if and only if \[
G(j\omega,\rho)+G^{*}(jw,\rho)\ge0,\forall\omega\in\mathbb{R},
\]
 which, according to \eqref{eq:closed-loop transfer 2}, is equivalent to 
\[
p_{N}^{e}(w)p_{D}^{e}(\omega,\rho)+p_{N}^{o}(w)p_{D}^{o}(\omega,\rho)\ge0,\forall\omega\in\mathbb{R}.
\]
 Therefore, there exists a controller in the set $\mathcal{C}$
that can feedback passivate the plant if and only if there exists
a vector $\rho\in\mathcal{P}$ and a scalar $\epsilon\ge0$ such that
\[
p_{N}^{e}(w)p_{D}^{e}(\omega,\rho)+p_{N}^{o}(w)p_{D}^{o}(\omega,\rho)-\epsilon\ge0,\forall\omega\in\mathbb{R},
\]
Since there is no gap between nonnegative polynomials and SOS polynomials when the polynomial is univariate, the above condition is  is equivalent to 
\[
p_{N}^{e}(w)p_{D}^{e}(\omega,\rho)+p_{N}^{o}(w)p_{D}^{o}(\omega,\rho)-\epsilon\text{ is SOS},
\]
which completes the proof. 
\endprf

Note that checking the feasibility of \eqref{eq:lmi cond 1} can be solved by a SDP. Specifically, the condition in \eqref{eq:lmi cond 1}
can be rewritten based on Section \ref{sos} as 
\[
M(\rho,\epsilon)+L(\alpha)\ge0
\]
where $M(\rho,\epsilon)$ is a matrix depending linearly on $(\rho,\epsilon)$
while $L(\alpha)$ a linear parametrization of the set defined in \eqref{L}. Moreover,
to take the constraint $\rho\in\mathcal{P}$ into account, the feasibility
problem in \eqref{eq:lmi cond 1} can be equivalently solved by checking the positivity
of $\epsilon^{*}$, which is the optimal solution of the following
SDP: 
\begin{equation}
\begin{array}{rcl}
\epsilon^{*} & = & \text{max}_{\rho,\alpha,\epsilon}\ \epsilon\\
 &  & \left\{ \begin{array}{c}
M(\rho,\epsilon)+L(\alpha)\ge0\\
\begin{pmatrix}\overline{\rho}_{i}-\rho_{i} & 0\\
0 & \rho_{i}-\underline{\rho}_{i}
\end{pmatrix}\ge0\\
i=1,\ldots,p
\end{array}\right.
\end{array}\label{eq:lmi}
\end{equation}

If the condition in \eqref{eq:lmi cond 1} is feasible, i.e., $\epsilon^*>0$, let
us further address the second part of Problem \ref{prob:2}. Consider the problem of desgining an optimal IFP controller. According to Lemma \ref{lem:relation} and  \eqref{eq:closed-loop transfer 2},
the problem can be equivalently rephrased in the following mathematical
form
\begin{equation}
\begin{array}{cl}
\underset{\rho\in\mathcal{P}}{\text{max}} & \nu\\
\text{s.t. } & \displaystyle \frac{p_{N}^{e}(w)p_{D}^{e}(\omega,\rho)+p_{N}^{o}(w)p_{D}^{o}(\omega,\rho)}{p_{D}^{e}(\omega,\rho)^{2}+p_{D}^{o}(\omega,\rho)^{2}}\ge\nu,  \forall\omega\in\mathbb{R}.
\end{array}\label{eq:max_IFP}
\end{equation}

\begin{thm}
\label{thm:CT_IFP}If the condition in \eqref{eq:lmi cond 1} is feasible, the maximum IFP index $\nu^{*}$ that can be achieved
by the feedback controller set $\mathcal{C}$ is given by $\nu^{*}=(\gamma^{*})^{2}$
with $\gamma^{*}$ defined as 
\begin{equation}
\begin{array}{l}
\gamma^{*}=\underset{\rho\in\mathcal{P},\gamma}{\text{max}}\ \gamma\ \text{s.t.}\\
\text{}\begin{pmatrix}\begin{array}{ccc}
p_{N}^{e}(w)p_{D}^{e}(\omega,\rho)+p_{N}^{o}(w)p_{D}^{o}(\omega,\rho) & \star & \star\\
\gamma p_{D}^{e}(\omega,\rho) & 1 & 0\\
\gamma p_{D}^{o}(\omega,\rho) & 0 & 1
\end{array}\end{pmatrix}\ \text{is SOS} \end{array}\label{eq:lmi cond _2}
\end{equation}
and the corresponding controller is given by the optimal solution
$\rho^{*}$. 
\end{thm}
\textbf{Proof.}
Suppose the condition in \eqref{eq:lmi cond 1} is feasible,
it follows that there exists $\bar{\rho}\in\mathcal{P}$ such that
\[
p_{N}^{e}(w)p_{D}^{e}(\omega,\bar{\rho})+p_{N}^{o}(w)p_{D}^{o}(\omega,\bar{\rho})\ge0,\forall \omega\in \mathbb{R}
\]
and hence,
\[
\frac{p_{N}^{e}(w)p_{D}^{e}(\omega,\bar{\rho})+p_{N}^{o}(w)p_{D}^{o}(\omega,\bar{\rho})}{p_{D}^{e}(\omega,\bar{\rho})^{2}+p_{D}^{o}(\omega,\bar{\rho})^{2}}\ge0, \forall \omega\in \mathbb{R}.
\]
Therefore, a lower bound of the optimal $\nu^{*}$ in \eqref{eq:max_IFP}
is zero. Next, let us observe that the constraint in \eqref{eq:max_IFP} can be rewritten as
\[
\begin{array}{l}
p_{N}^{e}(w)p_{D}^{e}(\omega,\rho)+p_{N}^{o}(w)p_{D}^{o}(\omega,\rho)-\nu p_{D}^{e}(\omega,\rho)^{2}\\-\nu p_{D}^{o}(\omega,\rho)^{2}
\ge0.
\end{array}
\]
 Since $\nu\ge0$ and by exploiting the Schur complement lemma, the
above inequality can be further equivalently rewritten as 
\[
\begin{pmatrix}\begin{array}{rcl}
p_{N}^{e}(w)p_{D}^{e}(\omega,\rho)+p_{N}^{o}(w)p_{D}^{o}(\omega,\rho) & \star & \star\\
\sqrt{\nu}p_{D}^{e}(\omega,\rho) & 1 & 0\\
 \sqrt{\nu}p_{D}^{o}(\omega,\rho)&0 & 1
\end{array}\end{pmatrix}\ge0.
\]
According to Theorem 4 in \cite{chesi2010lmi}, we have that a univariate matrix
polynomial is positive semidefinite if and only if it is SOS. Therefore, by replacing $\sqrt{\nu}$ with $\gamma$,
the optimization problem in \eqref{eq:max_IFP} can be equivalently
solved by \eqref{eq:lmi cond _2}, which completes the proof.
\endprf
\vspace{1mm}
\begin{rem}
Theorem \ref{thm:CT_IFP} provides a method via solving a convex optimization problem 
to design the controller in the  set $\mathcal{C}$ that  maximizes the  IFP index for the closed-loop system.  Particularly, the maximum $\gamma^{*}$ in the convex optimization problem \eqref{eq:lmi cond _2} can be obtained  by bisection
algorithm (i.e., at each step of the bisection algorithm, fix the
value of $\gamma$ and check the feasibility of \eqref{eq:lmi cond _2} ).  To check the feasibility of \eqref{eq:lmi cond _2} with fixed  value of $\gamma$, let us observe that the matrix in \eqref{eq:lmi cond _2}
depends linearly on the decision variables $\rho$, and the constraint
$\rho\in\mathcal{P}$ can be imposed by adding extra LMI constraints
as done in \eqref{eq:lmi}. Similar to the scalar polynomial case in \eqref{eq:lmi cond 1}, the condition for a matrix polynomial which depends on some decision variables linearly to be SOS polynomials can be solved equivalently via a SDP, as shown
in Section \ref{sos}. 
\end{rem}

Next, we consider the optimal OFP controller design. To this end, let us observe that the zeros of the closed-loop transfer function \eqref{eq:closed-loop transfer 1} have negative real part under Assumption \ref{assum:2} and the stable controller base $\bar{C}$. Therefore, the closed-loop system $G(s,\rho)$ is minimum phase system. Now, we are ready the present the following theorem. 

\begin{thm}
 The maximum OFP index $\xi^{*}$ that can be achieved
by the feedback controller set $\mathcal{C}$  is given by $\xi^{*}$ defined as 
\begin{equation}
\begin{array}{rl}
\xi^{*}=&\underset{\rho\in\mathcal{P}}{\text{max}} \;\; \xi\\
 & \text{s.t. } p_{N}^{e}(w)p_{D}^{e}(\omega,\rho)+p_{N}^{o}(w)p_{D}^{o}(\omega,\rho)-\\ & \;\;\;\;\;\;\xi p_{N}^{e}(\omega)^{2}-\xi p_{N}^{o}(\omega)^{2}\;\; \text{is SOS}.
\end{array}\label{eq:max_oFP}
\end{equation}
and the corresponding controller is given by the optimal solution
$\rho^{*}$. 
\end{thm}
\textbf{Proof.}
Since the closed-loop system $G(s,\rho)$ is minimum phase, its inverse exists. From \eqref{eq:CT_closed-loop jw}, we have that 
\[
G^{-1}(j\omega,\rho)=\frac{p_{D}^{e}(\omega,\rho)+jp_{D}^{o}(\omega,\rho)}{p_{N}^{e}(w)+jp_{N}^{o}(w)},
\]
and 
\[
\begin{array}{rl}
     & G^{-1}(j\omega,\rho)+[G^{-1}(j\omega,\rho)]^{*} \\
    = & \displaystyle \frac{2p_{D}^{e}(\omega,\rho)p_{N}^{e}(w)+2p_{D}^{o}(\omega,\rho)p_{N}^{o}(w)}{p_{N}^{e}(w)^2+p_{N}^{o}(w)^2}
\end{array}
\]
It follows from  Lemma \ref{lem:relation} that the maximum OFP index that can be reached is 
\[
\begin{array}{rl}
\xi^{*}=&\underset{\rho\in\mathcal{P}}{\text{max}} \;\; \xi\\
 & \text{s.t. } \displaystyle \frac{p_{D}^{e}(\omega,\rho)p_{N}^{e}(w)+p_{D}^{o}(\omega,\rho)p_{N}^{o}(w)}{p_{N}^{e}(w)^2+p_{N}^{o}(w)^2}\ge \xi, \forall \omega\in\mathbb{R},
 \end{array}
\]
 which can be rewritten into \eqref{eq:max_oFP}.
\endprf

Similar to the optimization problem \eqref{eq:lmi cond 1}, since the polynomial in \eqref{eq:max_oFP} depends linearly on decision variables $\rho$ and $\xi$, it can be solved by a SDP. 
\vspace{2mm}
\begin{rem}\label{rem:final}
An alternative approach to address directly Problem \ref{prob:2} without assuming that the set of $\mathcal{C}$ is stabilizing is
to solve the SDP presented in Theorem \ref{thm:CT_IFP}, and then
check the stability of the closed-loop system with the derived controller
$C(\rho^{*})$. See Example 2 in Section \ref{sec:example} for more details. 
\end{rem}

\subsubsection{DT case}

In the end, we consider Problem \ref{prob:2} for the discrete-time systems \eqref{eq:close-loop system_DT}.
In order to establish whether a given stable closed-loop system is
passive, we need to check the positivity of the real part of the transfer
function $G(z,\rho)$ over the complex unit circle $\{z\in\mathbb{C}:|z|=1\}$. 

Let $y\in\mathbb{R}$ be an auxiliary variable, and define the rational
function as $\phi:\mathbb{R}\rightarrow\mathbb{C}$ as $\phi(y)=\frac{1-y^{2}+j2y}{1+y^{2}}$. Note  that the complex unit circle $|z|=1$ is parameterized by the
variable $y\in\mathbb{R}$  (\cite{chesi2019stability}). Consequently, one has that 
\begin{equation}
 G(z,\rho)+G^{*}(z,\rho)\ge0,\forall|z|=1  
 \label{eq: frequency_condi_DT}
\end{equation}
is equivalent to 
\begin{equation}
 G(\phi(y),\rho)+G^{*}(\phi(y),\rho)\ge0,\forall y\in\mathbb{R}.   
\end{equation}

 Let us denote the numerator and denominator of the transfer function
in \eqref{eq:close-loop system_DT} as the polynomials $p_{N}(z)=\sum_{i=0}^{d_{N}}q_{i}z^{i}$
and $p_{D}(z,\rho)=\sum_{i=0}^{d_{D}}b_{i}(\rho)z^{i}$, respectively,
wherein the coefficients $b_{i}(\rho),i=1,\ldots,d_{D}$ depend linearly
on the vector variable $\rho$. By substituting $z=\phi(y)$, we have
\begin{equation}
\begin{array}{rcl}
p_{N}(\phi(y)) & = & \sum_{i=0}^{d_{N}}q_{i}\left(\frac{1-y^{2}+j2y}{1+y^{2}}\right)^{i}\\
p_{D}(\phi(y),\rho) & = & \sum_{i=0}^{d_{D}}b_{i}(\rho)\left(\frac{1-y^{2}+j2y}{1+y^{2}}\right)^{i}.
\end{array}
\end{equation}
By even-odd decomposition, it follows that $p_{N}(\phi(y))$ and $p_{D}(\phi(y),\rho)$ can be
expressed as 
\[
p_{N}(\phi(y))=\frac{p_{1}(y)+jp_{2}(y)}{(1+y^{2})^{d_{N}}}
\]
\[
p_{D}(\phi(y),\rho)=\frac{p_{3}(y,\rho)+jp_{4}(y,\rho)}{(1+y^{2})^{d_{D}}}
\]
where $p_{1},p_{2},p_{3},p_{4}$ are all real polynomials in $y$ with
coefficients of $p_{3},p_{4}$ depending linearly on $\rho$. Now it is ready to see
\[
G(\phi(y),\rho)=(1+y^{2})^{d_{D}-d_{N}}\frac{p_{1}(y)+jp_{2}(y)}{p_{3}(y,\rho)+jp_{4}(y,\rho)}
\]
which follows that
\begin{equation}
\begin{array}{rl}
   & G(\phi(y),\rho)+G^{*}(\phi(y),\rho)\\
  = & \displaystyle 2(1+y^{2})^{d_{D}-d_{N}}\frac{p_{1}(y)p_{3}(y,\rho)+p_{2}(y)p_{4}(y,\rho)}{p_{3}^{2}(y,\rho)+p_{4}^{2}(y,\rho)}.
\end{array}\label{eq:cloes-loop_DT}
\end{equation}

Since the given set of controllers $\mathcal{C}$ is stabilizing, it follows from Lemma \ref{lem:relation} and Definition \ref{def:Positive-realness} that the closed-loop system is passive if and only if the condition \eqref{eq: frequency_condi_DT} holds. Based on similar reasoning of Lemma \ref{lm:3}, we can obtain the following result.

\begin{lem}
\label{lem:DT_passivity}There exists a controller $C(z)$ in the
controller set $\mathcal{C}$ that can feedback passivate the plant
$G_{o}(z)$ if and only if there exists a vector $\rho\in\mathcal{P}$
and a scalar $\epsilon\ge0$ such that 
\begin{equation}
p_{1}(y)p_{3}(y,\rho)+p_{2}(y)p_{4}(y,\rho)-\epsilon\text{ is SOS}.\label{eq:DT_passivity}
\end{equation}
\end{lem}

Lemma \ref{lem:DT_passivity} provides, for the DT case,
a necessary and sufficient condition for determining the existence
of a controller $C(z)$ in the set $\mathcal{C}$ such that the closed-loop
system  \eqref{eq:close-loop system_DT} is  passive. Similar
to the CT case, this condition can be verified by solving a
SDP with the same form in \eqref{eq:lmi}. 

When the condition \eqref{eq:DT_passivity} is satisfied, the next
step is to determine the controller $C^{*}$ in the set $\mathcal{C}$
that can achieve the maximum IFP index $\nu^{*}$ for the closed-loop
system \eqref{eq:cloes-loop_DT}. 

\begin{cor}
If the condition in \eqref{eq:DT_passivity} is feasible, the maximum IFP index $\nu^{*}$ that can be achieved by the feedback controller
set $\mathcal{C}$ is given by $\nu^{*}=(\gamma^{*})^{2}$
with $\gamma^{*}$ defined as 
\begin{equation}
\begin{array}{l}
\gamma^{*}=\underset{\rho\in\mathcal{P},\gamma}{\text{max}}\ \gamma \ \text{s.t.}\\
\text{}\begin{pmatrix}\begin{array}{rcl}
\bar{p}(y,\rho) & \gamma p_{3}(y,\rho) & \gamma p_{4}(y,\rho)\\
\gamma p_{3}(y,\rho) & 1 & 0\\
 \gamma p_{4}(y,\rho)& 0 & 1
\end{array}\end{pmatrix}\ \text{is SOS}\\
\\
\end{array}\label{eq:IFP DT sdp}
\end{equation}
where 
\[
\bar{p}(y,\rho)=(1+y^{2})^{d_{D}-d_{N}}\left(p_{1}(y)p_{3}(y,\rho)+p_{2}(y)p_{4}(y,\rho)\right).
\]
\end{cor}
\vspace{2mm}
By taking the inverse of $G(\phi,\rho)$, it is obtained that
\begin{equation}
\begin{array}{rl}
   & G^{-1}(\phi(y),\rho)+[G^{-1}(\phi(y),\rho)]^{*}\\
  = & \displaystyle2(1+y^{2})^{d_{N}-d_{D}}\frac{p_{1}(y)p_{3}(y,\rho)+p_{2}(y)p_{4}(y,\rho)}{p_{1}^{2}(y)+p_{2}^{2}(y)}.
\end{array}
\end{equation}
\begin{cor}
 The maximum OFP index $\xi^{*}$ that can be achieved
by the feedback controller set $\mathcal{C}$ is given by $\xi^{*}$ defined as 
\begin{equation}
\begin{array}{rl}
\xi^{*}=&\underset{\rho\in\mathcal{P}}{\text{max}} \;\; \xi\\
\text{s.t. } &  (1+y^{2})^{d_{N}-d_{D}}\left(p_{1}(y)p_{3}(y,\rho)+p_{2}(y)p_{4}(y,\rho)\right)\\ & \;\;\;\;\;\;-\xi p_{1}^{2}(y)-\xi p_{2}^{2}(y)\;\; \text{is SOS}.
\end{array}\label{eq:max_OFP_DT}
\end{equation}
and the corresponding controller is given by the optimal solution
$\rho^{*}$. 
\end{cor}
\begin{rem}\label{rem:footnote}
It can be easily seen that the proposed methodology in this subsection can be used not only for a hyper-rectangle set $\mathcal{P}$ as defined in \eqref{eq:P set}, but also any convex  set $\mathcal{P}$. Indeed, this can be achieved by replacing the LMI the second constraint in \eqref{eq:lmi} with appropriate LMI corresponding to the set $\rho\in\mathcal{P}$.
\end{rem}

\section{Numerical Examples}\label{sec:example}

In this section, we provide two examples to illustrate the proposed 
methodologies. The computations are done by Matlab with toolbox SOSTOOLS 
and SeDuMi.
\subsection{Example 1}
Let us begin with considering a DT plant with transfer function $G_{0}(z)=\frac{z}{z-2}$,
and the controller set $\mathcal{C}$  described by $C(z,\rho)=\rho_{1}+\rho_{2}\frac{1}{z-0.5}$
with the parameter $\rho\in\mathcal{P}=[0.1,1]\times[1,2]$. Note
that the plant is unstable since its pole has magnitude larger than
1. The closed-loop system \eqref{eq:close-loop system_DT} is derived as
\[
G(z,\rho)=\frac{2z^{2}-z}{(2\rho_{1}+2)z^{2}+(2\rho_{2}-\rho_{1}-5)z+2}.
\]

The first problem is to establish whether the closed-loop system is
stable for all $\rho\in\mathcal{P}.$ To address this, we first compute
the modified Jury table  for the
denominator $p_N(z,\rho)$, and the first column of the table
is obtained as $f_{1}=2\rho_{1}+2,f_{2}=4\rho_{1}^{2}+8\rho_{1},f_{3}=12\rho_{1}^{4}+16\rho_{1}^{3}\rho_{2}+24\rho_{1}^{3}-16\rho_{1}^{2}\rho_{2}^{2}+80\rho_{1}^{2}\rho_{2}-36\rho_{1}^{2}.$
Next, we examine the positivity of these polynomials over the set $\rho\in\mathcal{P}$
based on Theorem \ref{thm:stability sos}. It is obvious that $f_{1}(\rho)>0$
and $f_{2}(\rho)>0$ for all $\rho\in\mathcal{P}$, so we just need
to solve the SOS program in \eqref{eq:stability_cond_2_2} for $i=3$.
By choosing the the degrees of the auxiliary polynomials $s_{31}(\rho),s_{32}(\rho)$
as 2, we find the optimal solution as $\theta^{*}=0.32$, which guarantees
the positivity of $f_{3}(\rho)$ over $\rho\in\mathcal{P}$. Therefore,
it can be concluded that the closed-loop system $G(z,\rho)$ is stable
for all $\rho\in\mathcal{P}$.
\vspace{1mm}

With this set of  stabilizing controllers, we further consider optimal IFP controller design in Problem \ref{prob:2}. The first step is to determine the existence
of controllers in the set $\mathcal{C}$ that can feedback passivate
the plant. This can be done by solving the SDP in \eqref{eq:DT_passivity}.
Specifically, by replacing $z$ with $\phi(y)=\frac{1-y^{2}+j2y}{1+y^{2}}$,
we have 
\[
 \begin{array}{l}
G(\phi(y),\rho)+G^{*}(\phi(y),\rho)=\displaystyle\frac{p_{1}(y)p_{3}(y,\rho)+p_{2}(y)p_{4}(y,\rho)}{p_{3}^{2}(y,\rho)+p_{4}^{2}(y,\rho)}\\
p_{1}(y)=3y^{4}-12y^{2}+1\\
p_{2}(y)=-10y^{3}+6y\\
p_{3}(y)=(3\rho_{1}-2\rho_{2}+9)y^{4}-(12\rho_{1}+8)y^{2}+\rho_{1}+2\rho_{2}-1\\
p_{4}(y)=(-10\rho_{1}+4\rho_{2}-18)y^{3}+(6\rho_{1}+4\rho_{2}-2)y.
\end{array}
\]

Then, we solve the SOS program in \eqref{eq:DT_passivity}, which is
converted to solving a SDP in the form of \eqref{eq:lmi}, and it
is obtained that the optimal solution of $\epsilon$ in \eqref{eq:lmi}
is positive. Therefore, it can be concluded that there exists a controller
in the set $\mathcal{C}$ that can feedback passivate the plant $G_{0}$.
The next step is to derive the controller $C^{*}$ in the set $\mathcal{C}$
that maximizes IFP $(\nu)$ for the closed-loop
system. This is accomplished by solving the SDP \eqref{eq:IFP DT sdp}
at each step of the bisection algorithm, which leads to the maximum
$\text{\ensuremath{\nu}}$ as $\nu^{*}=0.48$ with the optimal solution
$\rho_{1}^{*}=0.1,\rho_{2}^{*}=1.5$. 

\vspace{1mm}
To verify the resulting IFP index $\nu^{*}$, one can transform the
closed-loop transfer function $G(z,\rho^{*})$ to a state space system
$(A,B,C,D)$, and then exploits the necessary and sufficient LMI condition
for dissipativity to obtain the  IFP index for the closed-loop system. (See Lemma 2 in \cite{kottenstette2014relationships} for details) 
To be specific, the closed-loop system $G(z,\rho^{*})$ can be rewritten
as the state space system as follows
\[
\begin{array}{c}
A=\left(\begin{array}{cc}
0.955 & 0.91\\
1 & 0
\end{array}\right),B=\begin{pmatrix}1\\
0
\end{pmatrix},\\
C=\begin{pmatrix}0.413 & -0.826\end{pmatrix},D=0.91.
\end{array}
\]
It can be verified by Lemma 3 in \cite{kottenstette2014relationships} that the IFP index for this state space system is obtained as $0.48$ as expected.

\subsection{Example 2}
In this example, we consider a CT plant with the transfer function
$G_{0}(s)=\frac{(s+2)(s+3)}{(s-1)(s-2)}$, and the controller set
$\mathcal{C}$ is chosen to be the class of PI controllers, described
as $C(s,\rho)=\rho_{1}+\rho_{2}\frac{1}{s+1}$ with the parameter
$\rho\in\mathcal{P}=\left[0,1\right]\times\left[0,1\right]$.

The problem is to directly determine the controller $C^{*}$ in the set $\mathcal{C}$ that maximize the IFP index and the OFP index for the closed-loop system, respectively.
Let us observe that the plant is unstable since it has poles $\{1,2\}$.

First, by substituting $s=j\omega$, one can express the closed-loop
system \eqref{eq:CT_closed-loop jw} as 
\[
 \begin{array}{l}
G(j\omega,\rho) = \displaystyle \frac{p_{N}^{e}(w)+jp_{N}^{o}(w)}{p_{D}^{e}(\omega,\rho)+jp_{D}^{o}(\omega,\rho)}\\
p_{N}^{e}(w)  =  -6\omega^{2}+6\\
p_{N}^{o}(w)  =  -\omega^{3}+11\omega\\
p_{D}^{e}(\omega,\rho)  =  (2-6\rho_{1}-\rho_{2})\omega^{2}+(2+6\rho_{1}+6\rho_{2})\\
p_{D}^{o}(\omega,\rho)  =  (-1-\rho_{1})\omega^{3}+(-1+11\rho_{1}+5\rho_{2})\omega.
\end{array}
\]

Then, we solve the SOS progam in \eqref{eq:lmi cond 1}, which is
converted to solving the SDP \eqref{eq:lmi}, and it is obtained that
the optimal solution of $\epsilon$ in \eqref{eq:lmi} is positive.

To design the optimal IFP controller, we consider the optimization problem in \eqref{eq:max_IFP}.
By solving the SDP \eqref{eq:lmi cond _2} at each step of bisection
algorithm, we obtain that the maximum $\text{\ensuremath{\nu}}$ as
$\nu^{*}=0.658$ with the  solution $\rho_{1}^{*}=0.516,\rho_{2}^{*}=0.669$. To design the optimal OFP controller, we solve the optimization problem in \eqref{eq:max_oFP}, and obtain that the maximum $\xi$ as $\xi^{*}=0.542$ with the solution $\rho_1=\rho_2=1$. 

In the end, we need to check the stability of the closed-loop system \eqref{eq:closed-loop transfer 1}
with the derived $\rho^{*}$. For  both the optimal IFP controller $\rho_{1}^{*}=0.516,\rho_{2}^{*}=0.669$  and the optimal OFP controller $\rho_1=\rho_2=1$, the closed-loop system  $G(s,\rho^{*})=\frac{G_{0}(s)}{1+G_{0}(s)C(s,\rho^{*})}$  can be easily verified via Routh-Hurwitz stability criterion or
calculating the poles that the closed-loop system $G(s,\rho^{*})$
is stable. Therefore, based on Lemma \ref{lem:relation}, one has
that the maximum IFP index that the closed-loop system can achieve
is $\nu^{*}=0.658$, and the corresponding controller is $C(s,\rho^{*})=0.516+\frac{0.669}{s+1},$ and the maximum OFP index that the closed-loop system can achieve is $\xi^{*}=0.542$ and the corresponding controller is $
C(s,\rho^{*})=1+\frac{1}{s+1}$.

Similar to the previous example, the resulting IFP index $\nu^{*}$ and OFP index $\xi^{*}$
can be verified by transforming the closed-loop transfer function
$G(s,\rho^{*})$ to state space system $(A,B,C,D)$, and then exploit
the necessary and sufficient LMI condition for dissipativity (Lemma
2 in \cite{kottenstette2014relationships}) to obtain the  IFP or OFP index for the closed-loop system. 

\section{Conclusion}
This paper has considered feedback passivation of SISO LTI systems with linearly parameterized controller with the objective of maximizing the passivity level for the closed-loop systems.  First, we have  proposed a method to test whether a given set of controllers is stabilizing. Second, we have shown that given a set of stabilizing controllers,  the optimal controller in the sense  of maximum IFP or OFP index  can be obtained by solving a SDP. The proposed results also provide an alternative method without assuming the set of controllers to be stabilizing. Future work will consider extensions to the multi-input multi-output (MIMO) case.

\bibliography{su.bib}
\end{document}